\documentclass[twoside,fleqn]{article}
\usepackage{npb,axodraw,amsmath,alltt}

\newcommand{\FA}{\textsl{FeynArts}}
\newcommand{\mma}{\textsl{Mathematica}}
\newcommand{\FC}{\textsl{FormCalc}}
\newcommand{\NP}{\textsl{NumPrep}}
\newcommand{\FO}{\textsl{FORM}}
\newcommand{\LT}{\textsl{LoopTools}}
\newcommand{\MW}{M_{\rm W}}
\newcommand{\MZ}{M_{\rm Z}}
\newcommand{\lbrac}{\symbol{123}}
\newcommand{\rbrac}{\symbol{125}}
\newcommand{\uscore}{\symbol{95}}
\newcommand{\eg}{e.g.\ }
\newcommand{\ie}{i.e.\ }
\newcommand{\vp}{\vphantom{p}}
\newcommand{\ul}[1]{\underline{#1}}
\newcommand{\sq}{$^{\text{2}}$}

\newcommand{\cpc}[3]{{\sl Comp. Phys. Commun.} {\bf #1} (19#2) #3}
\newcommand{\fp}[3]{{\sl Fortschr. Phys.} {\bf #1} (19#2) #3}
\newcommand{\np}[3]{{\sl Nucl. Phys.} {\bf #1} (19#2) #3}

\newcommand{\pr}[3]{{\sl Phys. Rev.} {\bf #1} (19#2) #3}
\newcommand{\zp}[3]{{\sl Z. Phys.} {\bf #1} (19#2) #3}

\begin{document}

\title{Automatic Loop Calculations with \FA, \FC, and \LT}

\author{Thomas Hahn\address{%
	Institut f\"ur Theoretische Physik, Universit\"at Karlsruhe \\
	D--76128 Karlsruhe, Germany
	\hfill {\small KA--TP--7--2000}}}

\begin{abstract}
This article describes three \mma\ packages for the automatic calculation
of one-loop Feynman diagrams: the diagrams are generated with \FA,
algebraically simplified with \FC, and finally evaluated numerically using
the \LT\ package. The calculations are performed analytically as far as
possible, with results given in a form well suited for numerical
evaluation. The latter is straightforward with the utility programs
provided by \FC\ (\eg for translation into Fortran code) and the
implementations of the one-loop integrals in \LT. The programs are also 
equipped for calculations in supersymmetric models.
\end{abstract}

\maketitle


\section{Introduction}

The precision of experimental data achieved at present colliders has in
many cases reached or exceeded the per cent level. Obviously a comparable
accuracy on the theoretical side is needed in order to draw significant
conclusions from such precise measurements. For many observables this
means that a one-loop calculation is the lowest acceptable approximation.

Doing one-loop calculations by hand is laborious and error-prone and in
some cases simply impossible. So for some time already, software packages
have been developed to automate these calculations (\eg
\cite{MeBD91,xloops}). Incidentally, full automation is possible only up
to one loop since no algorithms generic enough for the computation of
arbitrary multi-loop Feynman diagrams are known at present. One remaining
obstacle is that the existing packages generally tackle only part of the
problem, and one still has to spend considerable effort adapting
conventions etc.\ to make them work together.

In this paper the three \mma\ packages \FA, \FC, and \LT\ are presented
which work hand in hand. The user has to supply only small driver programs
whose main purpose is to specify the necessary input parameters. This
makes the whole system very ``open'' in the sense that the results are
returned as \mma\ expressions which can easily be manipulated, \eg to
select or modify terms.

\begin{figure}
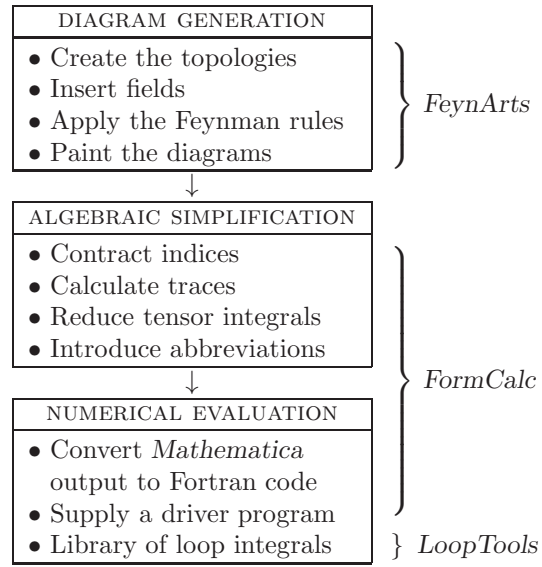

\begin{center}
\begin{tabular}{|l|l}
\cline{1-1}
\multicolumn{1}{|c|}{\sc diagram generation} \\
\cline{1-1}
$\bullet$ Create the topologies
          \vrule width 0pt height 2.5ex depth 0pt \\
$\bullet$ Insert fields \\
$\bullet$ Apply the Feynman rules \\
$\bullet$ Paint the diagrams
& \smash{\raise 4.3ex%
  \hbox{$\left.\vrule width 0pt depth 5ex height 0pt\right\}$ \FA}} \\
\cline{1-1}
\multicolumn{1}{c}{$\downarrow$} \\
\cline{1-1}
\multicolumn{1}{|c|}{\sc algebraic simplification} \\
\cline{1-1}
$\bullet$ Contract indices
          \vrule width 0pt height 2.5ex depth 0pt \\
$\bullet$ Calculate traces \\
$\bullet$ Reduce tensor integrals \\
$\bullet$ Introduce abbreviations
& \smash{\lower 2.7ex%
  \hbox{$\left.\vrule width 0pt depth 11.8ex height 0pt\right\}$ \FC}} \\
\cline{1-1}
\multicolumn{1}{c}{$\downarrow$} \\
\cline{1-1}
\multicolumn{1}{|c|}{\sc numerical evaluation} \\
\cline{1-1}
$\bullet$ Convert \mma
          \vrule width 0pt height 2.5ex depth 0pt \\
\hphantom{$\bullet$} output to Fortran code \\
$\bullet$ Supply a driver program \\
$\bullet$ Library of loop integrals
& $\left.\mathstrut\right\}\,$ \LT \\
\cline{1-1}
\end{tabular}
\end{center}
\vskip -10bp%
\caption{\label{fig:steps}%
Steps in a one-loop calculation and the distribution of tasks among the
programs \FA, \FC, and \LT.}
\end{figure}

Since one-loop calculations can range anywhere from a handful to several
hundreds of diagrams (particularly so in models with many particles like
the MSSM), speed is an issue, too. \FC, the program which does the
algebraic simplification, therefore uses \FO\ \cite{Ve91} for the
time-consuming parts of the calculation. Owing to \FO's speed, \FC\ can
process, for example, the 1000-odd one-loop diagrams of W--W scattering in
the Standard Model \cite{DeH98} in about 5 minutes on an ordinary Pentium
PC.

The following sections describe the main functions of each program.
Furthermore, the \FC\ package contains two sample calculations in the
electroweak Standard Model, $ZZ\to ZZ$ \cite{DeDH97} and $e^+e^-\to
\bar t\,t$ \cite{BeMH91}, which demonstrate how the programs are used
together.

\section{\FA}

\FA\ is a \mma\ package for the generation and visualization of Feynman
diagrams and amplitudes. The current version 2.2 is a much-expanded
version of \FA\ 1 \cite{KuBD91}. The two most important new features are
the generation of counter-term diagrams and the ability to deal with
supersymmetric theories (cf.\ Sect.\ \ref{sect:susy}).

\begin{figure}[t]
\begin{center}
\begin{small}
\unitlength=1bp%
\begin{picture}(190,275)(35,16)
\SetScale{.6}
\SetWidth{1.5}
\ArrowLine(150,411)(150,380)
\ArrowLine(150,335)(150,308)
\ArrowLine(150,251)(150,222)
\ArrowLine(150,175)(150,148)
\ArrowLine(150,101)(150,72)

\Line(200,50)(280,50)
\ArrowLine(279,50)(280,50)
\Text(180,32.8)[lb]{further}
\Text(180,28.8)[lt]{processing}

\ArrowArcn(210,260)(100,90,27)
\ArrowArc(210,300)(100,-90,-27)
\SetWidth{.5}

\GBox(62,410)(238,488){.9}
\Text(90,280.8)[b]{Find all distinct ways}
\Text(90,270.4)[b]{of connecting incoming}
\Text(90,260)[b]{and outgoing lines}
\Text(90,257.6)[t]{({\tt CreateTopologies})}
\GOval(150,360)(25,65)(0){1}
\Text(90,215.2)[]{Topologies}
\GBox(65,250)(235,310){.9}
\Text(90,174.4)[b]{Determine all allowed\vphantom{p}}
\Text(90,164)[b]{combinations of fields\vphantom{p}}
\Text(90,161.6)[t]{({\tt InsertFields})}
\GBox(248,250)(377,310){.9}
\Text(187,169.6)[b]{Draw the results\vphantom{p}}
\Text(187,165.6)[t]{({\tt Paint})}
\GOval(150,200)(25,65)(0){1}
\Text(90,119.2)[]{Diagrams}
\GBox(55,100)(245,150){.9}
\Text(90,76.8)[b]{Apply the Feynman rules}
\Text(90,72.8)[t]{({\tt CreateFeynAmp})}
\GOval(150,50)(25,65)(0){1}
\Text(90,29.6)[]{Amplitudes}
\end{picture}
\end{small}
\end{center}
\vskip -15bp%
\caption{\label{fig:feynarts}%
Flowchart for the generation of Feynman amplitudes with \FA.}
\end{figure}
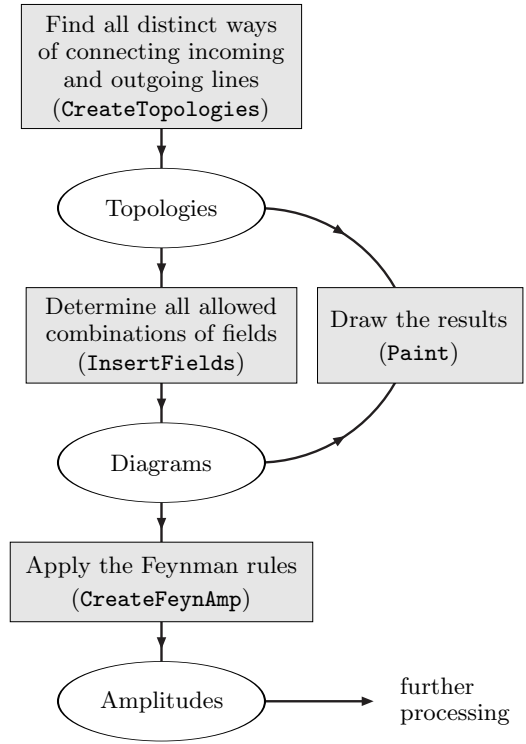

\FA\ works in three basic steps, sketched in Fig.~\ref{fig:feynarts}. The
first step is to create all different topologies for a given number of
loops and external legs. For example, to create all one-loop topologies
for a $1\to 2$ process, the following call to {\tt CreateTopologies} is
used:
\begin{verbatim}
  top = CreateTopologies[1, 1 -> 2]
\end{verbatim}

In the second step, the actual particles in the model have to be
distributed over the topologies in all allowed ways. E.g.\ the diagrams
for $Z\to b\bar b$ are produced with
\begin{alltt}
  ins = InsertFields[ top,
    V[2] -> \lbrac{}F[4,\,\lbrac{}3\rbrac{}], -F[4,\,\lbrac{}3\rbrac{}]\rbrac ]
\end{alltt}
where {\tt F[4,\,\lbrac 3\rbrac]} is the $b$-quark,
{\tt -F[4,\,\lbrac 3\rbrac]} its antiparticle, and \verb=V[2]= the $Z$
boson.

The fields, their propagators, and their couplings are defined in
a special file, the model file, which the user can supply or modify.
The following model files are included in \FA: the electroweak Standard
Model ({\tt SM.mod}) \cite{De93}, the same including QCD
({\tt SMQCD.mod}), and in the background-field formulation
({\tt SMbgf.mod}). These model files all include the full set of one-loop
counter terms. A model file for the MSSM is in preparation.

The diagrams can be drawn with \verb=Paint[ins]=, depending on the
options either on screen, or in a PostScript or \LaTeX\ file.

Finally, the analytic expressions for the diagrams are obtained by
\begin{verbatim}
  amp = CreateFeynAmp[ins]
\end{verbatim}

An important feature of \FA\ is that it distinguishes three levels of
fields:

-- {\em Generic level}, \eg the fermion {\tt F},

-- {\em Classes level}, \eg the down-type quark {\tt F[4]},

-- {\em Particles level}, \eg the $b$-quark {\tt F[4,\,\lbrac 3\rbrac]}.

\noindent
This is useful for two reasons:

The kinematic structure of a coupling is fixed once the generic fields are
specified. For example, all fermion--fermion--scalar couplings are of the
form
$$
C(F,F,S) = G_-\omega_- + G_+\omega_+
$$
where $\omega_\pm = (1\pm\gamma_5)/2$ are the chirality projectors. This
means that most algebraic simplifications like the tensor reduction need
to be carried out on the Generic-level amplitude only.

Furthermore, it is more economic to perform index summations (\eg over the
fermion-generation index) in a loop over Classes-level amplitudes instead
of generating many Particles-level diagrams.

\section{\FC}

\FA\ produces very symbolic output which cannot straightforwardly be
implemented in a numerical program. Its evaluation proceeds instead in two
steps: first, algebraic simplification in \mma; then, translation into a
Fortran program which computes the squared matrix element and from this
the desired quantities (cross-sections, decay rates, asymmetries, etc.).

\subsection{Algebraic simplification}

The symbolic expressions for the diagrams are simplified algebraically
with \FC\ which returns the results in a form well suited for numerical
evaluation. Specifically, \FC\ performs the following simplifications:

-- indices are contracted as far as possible,

-- fermion traces are evaluated,

-- open fermion chains are simplified using the \\
\indent\hphantom{--} Dirac equation,

-- colour structures are simplified using the \\
\indent\hphantom{--} SU($N$) algebra,

-- the tensor reduction is done,

-- the results are partially factored,

-- abbreviations are introduced.

\begin{figure}
\unitlength=1bp%
\begin{picture}(212,210)(-30,0)
\SetScale{.7}
\GBox(0,165)(260,300){.9}
\BBox(10,225)(250,290)
\Text(91,193.2)[]{\large\mma}
\Text(32,172)[br]{{\sc pro:}\vp}
\Text(35.5,172)[bl]{user friendly\vp}
\Text(32,160.3)[br]{{\sc con:}\vp}
\Text(35.5,160.3)[bl]{slow on large expressions\vp}

\BBox(10,175)(95,215)
\Text(36.75,141.4)[]{\FA}
\Text(36.75,129.5)[]{amplitudes}

\BBox(165,175)(250,215)
\Text(145.25,141.4)[]{\FC\vp}
\Text(145.25,129.5)[]{results\vp}

\GBox(0,0)(260,110){.9}
\BBox(10,10)(250,76)
\Text(91,42.7)[]{\large\FO}
\Text(32,21.7)[br]{{\sc pro:}\vp}
\Text(35.5,21.7)[bl]{very fast for certain operations\vp}
\Text(32,9.8)[br]{{\sc con:}\vp}
\Text(35.5,9.8)[bl]{not so user friendly\vp}

\DashLine(-40,156)(260,156){4}
\DashLine(-40,300)(-3,300){4}
\DashLine(-40,0)(-3,0){4}

\CBox(95,192)(111,198){White}{White}
\CBox(105,86)(111,198){White}{White}
\CBox(149,192)(170,198){White}{White}
\CBox(149,86)(155,198){White}{White}
\CBox(105,86)(155,92){White}{White}

\Line(90,198)(111,198)
\Line(111,198)(111,92)
\Line(111,92)(149,92)
\Line(149,92)(149,198)
\Line(149,198)(170,198)
\Line(170,198)(173,195)
\Line(173,195)(170,192)
\Line(170,192)(155,192)
\Line(155,192)(155,86)
\Line(155,86)(105,86)
\Line(105,86)(105,192)
\Line(105,192)(90,192)
\Line(90,192)(93,195)
\Line(93,195)(90,198)

\BBox(61,128)(124,146)
\Text(65.1,95.2)[]{\small input file}

\BBox(136,128)(199,146)
\Text(116.9,95.2)[]{\small\sl MathLink\vp}

\SetWidth{1.5}

\Line(-20,255)(-20,296)
\ArrowLine(-20,295)(-20,296)
\Line(-20,202)(-20,162)
\ArrowLine(-20,163)(-20,160)

\Text(-5,170.8)[r]{\FC\vp}
\Text(-5,158.9)[r]{user\vp}
\Text(-5,148.4)[r]{interface\vp}

\Line(-20,104)(-20,150)
\ArrowLine(-20,149)(-20,152)
\Line(-20,52)(-20,6)
\ArrowLine(-20,7)(-20,4)

\Text(-5,65.8)[r]{internal\vp}
\Text(-5,54.6)[r]{\FC\vp}
\Text(-5,43.4)[r]{functions\vp}
\end{picture}
\vskip -10bp%
\caption{\label{fig:fc}The interplay between \mma\ and \FO\ in \FC.}
\end{figure}
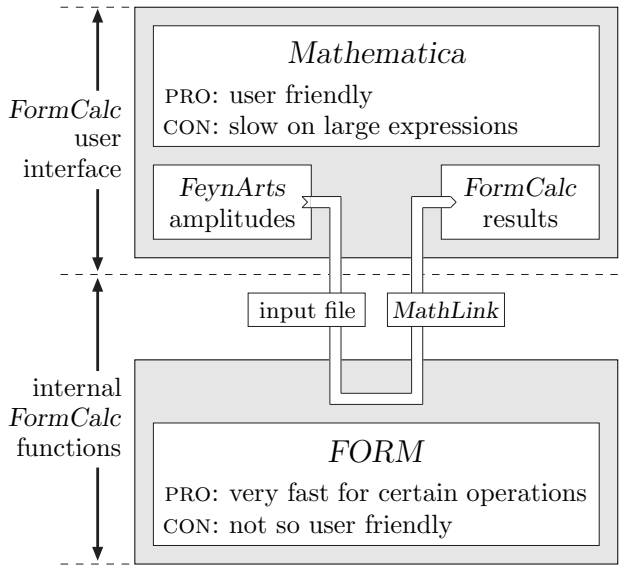

\noindent
The internal structure of \FC\ is simple: it prepares the symbolic
expressions of the diagrams in an input file for \FO, runs \FO, and
retrieves the results via the {\sl MathLink} interface (see
Fig.~\ref{fig:fc}). This is done completely without intervention by the
user, \ie the user does not see the \FO\ code. \FC\ thus combines the
speed of \FO\ with the powerful instruction set of \mma\ and the latter
greatly facilitates further processing of the results.

The main function in \FC\ is {\tt OneLoop} (the name is not strictly
correct since it works also with tree graphs). It is used like this:
\begin{verbatim}
  << FormCalc`
  amps = << myamps.m
  result = OneLoop[amps]
\end{verbatim}
where it is assumed that the file {\tt myamps.m} contains amplitudes
generated by \FA. Note that {\tt OneLoop} needs no declarations of the
kinematics of the underlying process; it uses the information \FA\ hands
down.

Even more comprehensive than {\tt OneLoop}, the function {\tt ProcessFile}
can process entire files. It collects the diagrams into blocks such that
index summations (\eg over fermion generations) can later be carried out
easily, \ie only diagrams which are summed over the same indices are put
in one block. This nicely complements the generation of Classes-level
diagrams in \FA, which leaves the index summations to the numerical
evaluation in order to reduce the number of diagrams. {\tt ProcessFile}
is invoked \eg as
\begin{verbatim}
  ProcessFile["vertex.amp", "vertex"]
\end{verbatim}
which reads the \FA\ amplitudes from the input file {\tt vertex.amp} and
produces output files of the form {\tt vertex{\it id}.m}, where {\it id}
is some identifier for a particular block.

The output of {\tt OneLoop} or {\tt ProcessFile} is in general a linear
combination of loop integrals with prefactors that contain model
parameters, kinematic variables, and abbreviations introduced by \FC. Such
abbreviations are introduced for spinor chains, scalar products of
vectors, and epsilon tensors contracted with vectors. A term in the
output could for instance look like
\begin{alltt}
C0i[cc0, MW2, MW2, S, MW2, MZ2, MW2] *
  ( -4 Alfa2 CW2 MW2/SW2 S AbbSum16 +
    32 Alfa2 CW2/SW2 S\sq AbbSum28 +
    4 Alfa2 CW2/SW2 S\sq AbbSum30 -
    8 Alfa2 CW2/SW2 S\sq AbbSum7 +
    Alfa2 CW2/SW2 S\,(T\,-\,U) Abb1 +
    8 Alfa2 CW2/SW2 S\,(T\,-\,U) AbbSum29 )
\end{alltt}
The first line represents the one-loop integral
$C_0(\MW^2, \MW^2, s, \MW^2, \MZ^2, \MW^2)$, which is multiplied with a
linear combination of abbreviations like {\tt Abb1} or {\tt AbbSum29} with
certain coefficients. These coefficients contain kinematical variables
like the Mandelstam variables {\tt S}, {\tt T}, and {\tt U} and model
parameters, \eg ${\tt Alfa2} = \alpha^2$.

The automatic introduction of abbreviations is a very important feature
which can drastically reduce the size of an amplitude, particularly so
because the abbreviations are nested in three levels. Here is an example:
\begin{center}
\begin{picture}(200,60)(5,10)
\Text(0,60)[bl]{\tt ~AbbSum29 = Abb2 + Abb22 + Abb23 + Abb3}
\EBox(97,56)(128,70)
\Line(97,56)(45,46)
\Line(45,46)(45,34)
\Line(128,56)(180,46)
\Line(180,46)(180,34)

\Text(113,36)[b]{\tt Abb22 = Pair1 Pair3 Pair6}
\EBox(118,32)(149,46)
\Line(118,32)(70,22)
\Line(70,22)(70,10)
\Line(149,32)(197,22)
\Line(197,22)(197,10)

\Text(134,11)[b]{\tt Pair3 = Pair[e[3],\,k[1]]}
\end{picture}
\end{center}
The definitions of the abbreviations can be retrieved by
{\tt Abbreviations[]} which returns a list of rules such that
{\tt result\,\,//.\,Abbreviations[]} gives the full, unabbreviated
expression.

\subsection{Translation to Fortran code}

For numerical evaluation, the \mma\ expressions produced by \FC\ need to
be translated into a Fortran program. (They could, in principle, be
evaluated in \mma\ directly, but this becomes rather slow for large
amplitudes.) The translation is done by the program \NP, which is part of
the \FC\ package. The philosophy of \NP\ is that the user should not have
to modify the generated code. This means that the code has to be
encapsulated (\ie no loose ends the user has to bother with), and that all
necessary subsidiary files (include files, makefile) have to be produced,
too.

From the point of view of the Fortran programmer who wants to use the
generated code in his program, the output of \NP\ is a single subroutine
called
\begin{center}
{\tt squared\uscore me($k_1,\dots,k_N,\,
  \varepsilon_1,\dots,\varepsilon_N,\,
  \lambda_1,\dots,\lambda_N$)}
\end{center}
which takes as input the external momenta, polarization vectors, and
helicities, and returns the squared matrix element. To obtain actual
numerical results from the generated code, one needs in addition a driver
program whose task is to initialize the model parameters, set up the
kinematics, invoke the {\tt squared\uscore me} subroutine, perform
necessary phase-space integrations, and finally write the results to a
file. A sample driver program for $2\to 2$ processes is included in \FC.

Finally, the generated code has to be linked with the \LT\ library which
provides the one-loop functions.

\section{\LT}

\LT\ supplies the actual numerical implementations of the one-loop
integrals needed for programs made from the \FC\ output. It is based on
the reliable package {\sl FF} \cite{vOV90} and provides in addition to the
scalar integrals of {\sl FF} also the tensor coefficients in the
conventions of \cite{De93}. \LT\ offers three interfaces: Fortran, C++,
and \mma, so most programming tastes should be served.

Using the \LT\ functions in Fortran and C++ is very similar. In Fortran
it is necessary to include the file {\tt looptools.h} in every function or
subroutine (for the common blocks). In C++, {\tt clooptools.h} must be
included once. Before using any \LT\ function, {\tt ffini} must be called
and at the end of the calculation {\tt ffexi} may be called to obtain a
summary of errors. It is of course possible to change parameters like the
scale $\mu$ from dimensional regularization; this is described in detail
in the manual \cite{LTGuide}.

A very simple Fortran program would for instance be
\begin{verbatim}
       program simple
  #include "looptools.h"
       call ffini
       print *, B0(1000D0,50D0,80D0)
       call ffexi
       end
\end{verbatim}
The C++-version of this program is
\begin{verbatim}
  #include "clooptools.h"
\end{verbatim}
\begin{verbatim}
  main()
  {
    ffini();
    cout << B0(1000.,50.,80.) << "\n";
    ffexi();
  }
\end{verbatim}
The \mma\ interface is even simpler to use:
\begin{verbatim}
  In[1]:= Install["LoopTools"]
\end{verbatim}
\begin{verbatim}
  In[2]:= B0[1000, 50, 80]
\end{verbatim}
\begin{verbatim}
  Out[2]= -4.40593 + 2.70414 I
\end{verbatim}

\section{Calculations in Supersymmetric Models}
\label{sect:susy}

Special emphasis has been placed on the possibility to do calculations
in supersymmetric models with \FA\ and \FC. In particular the following
two fundamental problems become relevant in supersymmetric theories:

\smallskip

\ul{\sc Problem 1:} SUSY theories in general contain Majorana fermions and
hence fermion-number-violating couplings (\eg quark--squark--gluino). The
textbook prescription of ordering the Dirac matrices opposite to their
occurrence along the arrows on fermionic lines obviously breaks down in
this case since one cannot define a fermion-number flow. (Put differently,
Majorana-fermion lines have no arrow.)

\ul{\sc Solution:} \FA\ uses the ``flipping-rule'' algorithm
\cite{DeEHK92}: instead of traversing the fermion lines along the
fermion-number flow imposed from the outside, it {\it chooses} a
particular direction for each fermion chain. If it later turns out that,
for a Dirac fermion, the chosen direction is opposite to the actual
fermion flow, it applies a so-called flipping rule.

\smallskip

\ul{\sc Problem 2:} Dimensional regularization, the default regularization
scheme employed by \FC, is known to break supersymmetry \cite{CJN80}.

\ul{\sc Solution:} \FC\ has two regularization schemes built in which are
chosen with the variable {\tt\$Dimension}. The default is {\tt\$Dimension
= D} which corresponds to dimensional regularization. Putting
{\tt\$Dimension = 4} switches to constrained differential renormalization
\cite{techniques}. The latter technique is equivalent at the one-loop
level to regularization by dimensional reduction \cite{HaP98} and is hence
suited for calculations in SUSY models.

\section{Requirements and Availability}

All three packages require \mma\ 2.2 or above; \FC\ needs in addition
\FO, preferably version 2 or above; \LT\ needs a Fortran compiler and
{\tt gcc}/{\tt g++}.

The packages should compile and run without change on any Unix
platform. They are specifically known to work under DEC Unix, HP-UX,
Linux, Solaris, and AIX. A comprehensive manual which explains
installation and usage is included in each package. All three packages 
are open-source programs and stand under the GNU library general public
license. They are available from

\quad{\tt http://www.feynarts.de}

\quad{\tt http://www.feynarts.de/formcalc}

\quad{\tt http://www.feynarts.de/looptools}

\end{document}